\documentclass[12pt]{article}
\usepackage{latexsym}
\usepackage{epsfig,amssymb,euscript,slashed}
\usepackage{amsmath}
\usepackage{array,calc,epsfig}
\usepackage[linktocpage]{hyperref}
\usepackage{pstricks}
\usepackage{bbm}
\usepackage{fancybox}

\def\be{\begin{equation}}
\def\ee{\end{equation}}
\def\bseq{\begin{subequations}}
\def\eseq{\end{subequations}}

\def\bea{\begin{eqnarray}}
\def\eea{\end{eqnarray}}

\def\bseq{\begin{subequations}}
\def\eseq{\end{subequations}}

\numberwithin{equation}{section} 
\usepackage[]{graphicx}

\def\d {{\rm d}}
\def\cala         {{\cal A}}

\def\calb         {{\cal B}}
\def\calc         {{\cal C}}

\def\calf         {{\cal F}}
\def\calg         {{\cal G}}

\def\calh         {{\cal H}}
\def\cali         {{\cal I}}
\def\calj         {{\cal J}}

\def\call         {{\cal L}}
\def\calm         {{\cal M}}
\def\caln         {{\cal N}}
\def\calo         {{\cal O}}

\def\calt         {{\cal T}}

\def\del          {\partial}
\def\delbar       {\bar\partial}
\def\ii           {{\rm i}}

\def\Re           {{\rm Re\hskip0.1em}}
\def\Im           {{\rm Im\hskip0.1em}}

\def\sqr#1#2{{\vcenter{\vbox{\hrule height.#2pt
 \hbox{\vrule width.#2pt height#1pt \kern#1pt \vrule width.#2pt}\hrule
 height.#2pt}}}}

\def\d{\text{d}}

\def\slashchar#1{\setbox0=\hbox{$#1$}           
\dimen0=\wd0                                 
\setbox1=\hbox{/} \dimen1=\wd1               
\ifdim\dimen0>\dimen1                        
\rlap{\hbox to \dimen0{\hfil/\hfil}}      
#1                                        
\else                                        
\rlap{\hbox to \dimen1{\hfil$#1$\hfil}}   
/                                         
\fi}

\begin{document}
\font\cmss=cmss10 \font\cmsss=cmss10 at 7pt

\begin{flushright}{\scriptsize DFPD-2016/TH/15 
}
\end{flushright}
\hfill
\vspace{18pt}
\begin{center}
{\Large \textbf{Warped K\"ahler potentials and fluxes}}
\end{center}

\vspace{6pt}
\begin{center}
{\textsl{Luca Martucci}}

\vspace{1cm}
\textit{Dipartimento di Fisica ed Astronomia ``Galileo Galilei",  Universit\`a di Padova\\
\& I.N.F.N. Sezione di Padova,
Via Marzolo 8, 35131 Padova, Italy} \\  \vspace{6pt}
\end{center}


\vspace{12pt}

\begin{center}
\textbf{Abstract}

\end{center}

\vspace{4pt} {\small
\noindent  The four-dimensional effective theory for type IIB warped flux compactifications proposed in \cite{Martucci:2014ska} is completed by   taking into account the  backreaction of the K\"ahler moduli on the three-form fluxes. The only required modification  consists in a flux-dependent contribution to the chiral fields parametrising the K\"ahler moduli. The resulting supersymmetric effective theory satisfies the no-scale condition and consistently combines previous partial results present in the literature. Similar results hold for M-theory warped compactifications on Calabi-Yau fourfolds, whose effective field theory and K\"ahler potential  are also discussed. 
\noindent }

\vspace{1cm}


\thispagestyle{empty}


\newpage

\setcounter{footnote}{0}

\tableofcontents
%

\section{Introduction}

Type IIB/F-theory  flux compactifications play a prominent role in string phenomenology, see for instance \cite{Denef:2008wq,Weigand:2010wm,Maharana:2012tu}  for reviews. One important aspect of such compactifications is that fluxes and branes can backreact on the geometry generating a non-trivial warp factor, which  can induce hierarchies in the energy scales  perceived  by the four-dimensional observers located at different points of the internal space. Nevertheless, the warp-factor is often assumed to be constant for technical simplicity, since this allows one to  borrow several of the well established results on unwarped Calabi-Yau compactifications. It is then important to formulate  effective four-dimensional  theories that go beyond the limits of the constant warping approximation and consistently combine  the effects of warping, fluxes and branes.

In particular, the non-trivial warp factor is expected to affect the K\"ahler potential and the kinetic terms of the effective theory, which set the strength of  physical masses and couplings, whose holomorphic structure is encoded in the F-terms. 
 In \cite{Martucci:2014ska}, following the ideas of \cite{eff1,eff2}, a derivation of the low-energy effective theory for the moduli sector of type IIB warped flux compactifications \cite{Grana:2000jj,Gubser:2000vg,GKP} has been proposed.\footnote{Other works on the effective theory of warped type IIB/M-theory  flux compactifications include \cite{Dasgupta:1999ss,DeWolfe:2002nn,Giddings:2005ff,Burgess:2006mn,Shiu:2008ry,Douglas:2008jx,Frey:2008xw,Marchesano:2008rg,Chen:2009zi,Underwood:2010pm,Marchesano:2010bs,Grimm:2012rg,Frey:2013bha,Grimm:2014efa,Grimm:2015mua,Cownden:2016hpf}.} The derivation exploits a combination of arguments based on the structure of the ten-dimensional vacua and the expected supersymmetric properties of the effective four-dimensional theory,  taking advantage of its  superconformal formulation at an intermediate step. Assuming that complex structure and 7-brane moduli  are stabilised by fluxes, the resulting K\"ahler potential for the remaining massless moduli has a  surprisingly simple form in terms of the universal modulus $a$ characterising this class of vacua:
\be\label{K0}
K=-3\log a \ .
\ee
The universal modulus implicitly depends on the chiral coordinates parametrising the complete set of moduli and such implicit dependence codifies the non-triviality of $K$.

The crucial step is then the determination of the appropriate complex/chiral coordinates on the moduli space. For instance, the  moduli describing the positions of the mobile D3-branes inherit the complex structure of the internal complex space while the two-form potentials naturally combine into  $C_2-\tau B_2$, which provides the appropriate complex parametrisation of the corresponding moduli. On the other hand,  the determination of the chiral coordinates $\rho_a$ parametrising the  K\"ahler and Ramond-Ramond (R-R) $C_4$ moduli is less obvious and in \cite{Martucci:2014ska} they were identified by means of supersymmetric probe Euclidean D3-branes.  The  resulting effective Lagrangian is directly affected by the warping   produced by the three-form fluxes and by the mobile D3-branes and consistently satisfies the no-scale condition  \cite{Cremmer:1983bf,Ellis:1983sf} in a non-obvious way. 

In the derivation  of \cite{Martucci:2014ska} the three-form fluxes were considered as fixed quantities specified by certain Dolbeault cohomology classes, which stabilise the axion-dilaton and complex structure moduli and  backreact on the geometry by warping it, but otherwise decouple from the low-energy dynamics. This assumption allows one to derive the effect of the three-form fluxes due to their induced D3-charge but, in fact, does not capture the complete potential impact of the fluxes on the low-energy physics. In order to understand this point, let us focus for simplicity on the vacua with constant axion-dilaton $\tau$ and recall that supersymmetry requires the three-form flux  $G_3\equiv F_3-\tau H_3$ to define a class in the $H^{2,1}(X)$ cohomology group of the internal space $X$, which clearly does not depend on the K\"ahler moduli.  On the other hand,  $G_3$ is required to be primitive, that is to satisfy $J\wedge G_3=0$, where $J$ is the K\"ahler form. Such primitivity condition is trivial at the cohomological level and  can be always satisfied for generic (strictly SU(3)-holonomy) compactifications, but it nevertheless introduces in  $G_3$ a  hidden dependence on the K\"ahler moduli. This  dependence was neglected in  \cite{Martucci:2014ska} but, as we will see, its inclusion potentially affects the low-energy  theory. 

In this paper it is shown how such hidden dependence can be taken into account by making a minimal modification of the chiral coordinates $\rho_a$ proposed in  \cite{Martucci:2014ska} while keeping the simple form (\ref{K0}) of the K\"ahler potential unchanged.   This modification is sensible even if one allows for a supersymmetry breaking flux component $G^{0,3}\neq 0$  and, consistently, the K\"ahler potential still satisfies the no-scale condition. The resulting effective Lagrangian is almost identical to the one obtained  in  \cite{Martucci:2014ska}. The only difference is in a $G_3$-dependent contribution to the kinetic metric for the $\rho_a$ moduli. Remarkably, the presence of the same contribution has already been  pointed out  in \cite{Frey:2013bha}, following an approach which is somewhat orthogonal to the procedure adopted here and in \cite{Martucci:2014ska}. Namely,  the authors of \cite{Frey:2013bha} performed a direct and careful dimensional reduction which focuses on the $C_4$-moduli and keeps the (non-universal) K\"ahler moduli fixed -- see also the very recent \cite{Cownden:2016hpf} for a  similar derivation including mobile D3-branes. Hence, the results of the present paper provide a  manifestly supersymmetric completion of the results of \cite{Frey:2013bha,Cownden:2016hpf} and  resolve  the apparent conflict between the approaches adopted in \cite{Frey:2013bha,Cownden:2016hpf} and in \cite{eff1,eff2,Martucci:2014ska}. 

In section \ref{sec:preliminaries}, the  above-mentioned $G_3$-induced  modifications of the effective theory are derived and discussed for the subclass of type IIB compactifications on warped Calabi-Yau spaces with constant axion-dilaton and no seven-branes. A purely type IIB extension of the results to F-theory models, although in principle straightforward, is complicated by the non-trivial transformation properties of $G_3$ under the SL$(2,\mathbb{Z})$ duality group and by its entanglement with the  fluxes supported on 7-branes.  
On the other hand,  the F-theory generalisation of the results of section \ref{sec:preliminaries} admit a simple formulation in the dual M-theory framework \cite{Vafa:1996xn},   in which   the type IIB bulk and seven-brane fluxes are unified in the background $G_4$-flux. This is shown  in section \ref{sec:Ftheory}, which considers M-theory warped flux compactifications to three-dimensions on general (non-necessarily elliptically fibered) Calabi-Yau four-folds  \cite{Becker:1996gj,Becker:2001pm}.  In section \ref{sec:Ftheory} we show how the logic followed in  \cite{Martucci:2014ska} and in section \ref{sec:preliminaries} of the present paper can be readily adapted to the M-theory framework, leading to a simple K\"ahler potential similar to (\ref{K0}) and to chiral coordinates that allow for the incorporation  of warping, mobile M2-branes and fluxes into the effective theory. In particular, we will see that the resulting effective Lagrangian contains  manifestly $G_4$ dependent terms, analogous to the type IIB $G_3$ dependent terms discussed above.


\section{Effective theory of warped IIB models}
\label{sec:preliminaries}

Let us review the structure of   IIB/F-theory warped flux compactifications \cite{Grana:2000jj,Gubser:2000vg,GKP}, mostly  following the notation of \cite{Martucci:2014ska}.  These vacua have an Einstein frame metric of the form
\be\label{10dmetric}
\d s^2_{10}=\ell_{\rm s}^2\,e^{2A}|\Psi|^2\d s^2_4+\ell_{\rm s}^2\, e^{-2A}\d s^2_{X}
\ee
where $\ell_s\equiv 2\pi \sqrt{\alpha'}$ is the string length scale, $\d s^2_4$ is the external four-dimensional Minkowski metric, $\d s^2_{X}$ is the metric of the internal K\"ahler space $X$ and $e^{A}$ represents the warp factor which varies along the internal directions $y^m$, $m=1,\ldots,6$. $\Psi$ is the conformal compensator, which has dimension of a mass and is eventually fixed in order to obtain a canonically normalised four-dimensional supergravity  in the Einstein frame.  Note that we have introduced the $\ell_{\rm s}^2$ factor in order to work with natural units along the internal space. 

The warp factor is sourced by the D3-charges present in the internal space and determined by the equation
\be\label{warpeq}
\Delta_X e^{-4A}=\frac{1}{\ell^4_{\rm s}}*_X Q_6
\ee
with $Q_6$ representing the D3-charge density
\be\label{D3charge}
Q_6=\ell^4_{\rm s}\sum_{I\in\text{D3's}}\delta_I^6-\frac{\ii}{2\Im\tau}G_3\wedge\bar G_3 +Q^{\rm nd}_6
\ee
where $\delta_I^6$ are delta-like 6-forms associated with the mobile D3-branes, while $Q^{\rm nd}_6$ is some additional contribution to the D3-charge, for instance due to O3-planes or curvature corrections on 7-branes, which we assume non-dynamical. The internal part of the self-dual five-form flux $F_5$ is completely fixed in terms of the warping, $F_5^{\rm int}=\ell_{\rm s}^4*_X\d e^{-4A}$, and the associated tadpole condition reads 
\be
\int_XQ_6=0\ .
\ee
The general solution of (\ref{warpeq}) can be written in the form
\be\label{warpsplit}
e^{-4A}=a+e^{-4A_0}
\ee
where $a$ is the  {\em universal modulus}   and $e^{-4A_0(y)}$ is the particular solution of (\ref{warpeq}) fixed by the condition
\be\label{A0cond}
\int_X e^{-4A_0}\d\text{vol}_X=0\ .
\ee
Notice that $e^{-4A_0(y)}$ can  also assume negative values very close to sources with negative tension, where the supergravity approximation breaks down. 

The internal K\"ahler metric $\d s^2_{X}$ has fixed (dimensionless) volume ${\rm v}_0$:
\be\label{volcond}
\int_X\d\text{vol}_X=\frac1{3!}\int_X J\wedge J\wedge J= {\rm v}_0
\ee
where $J$ is the K\"ahler form.\footnote{The notation of \cite{Martucci:2014ska} has been slightly simplified. In particular  $a$, $e^{-4A_0}$, $\d s^2_{X}$ and $J$ here correspond to $\hat a$, $e^{-4\hat A_0}$, $\d s^2_{X,0}$ and $J_0$ there.}
This does not imply a loss of any degree of freedom, since the overall breathing mode is parametrised by the universal modulus $a$. The remaining {\em non-universal} K\"ahler moduli $v^a$, $a=1,\ldots,h^{1,1}(X)-1$, can be identified with the coefficients of the expansion
\be\label{Jexp}
J=v^a\omega_a
\ee
where $\omega_a$ are integral harmonic $(1,1)$-forms which provide a basis of $H^2(X;\mathbb{Z})$.  Because of (\ref{volcond}), these non-universal K\"ahler moduli 
satisfy the constraint
\be\label{cc}
\frac1{3!}\cali_{abc}v^av^bv^c={\rm v}_0
\ee
where we have introduced the  intersection numbers
\be
\cali_{abc}\equiv \int_X\omega_a\wedge \omega_b\wedge\omega_c\ .
\ee

In the following a key role is played by the three-form flux
\be
G_3\equiv F_3-\tau H_3
\ee 
which, away from fluxed 7-branes, must satisfy the Bianchi identities $\d F_3=\d H_3=0$. 
 In order to solve the equations of motion, $G_3$ must be imaginary-self-dual ($*_XG_3=\ii G_3$) that is, it must have vanishing components  $G^{1,2}=G^{3,0}=0$ and be primitive:
\be\label{primitive}
J\wedge G_3=0
\ee  
This condition is accompanied by an analogous primitivity condition on the 7-branes fluxes. 
Furthermore, the $(2,1)$ and primitive component $G^{2,1}_{\rm P}$ preserves four-dimensional $\caln=1$ supersymmetry, while a non vanishing  $(0,3)$ component $G^{0,3}$ breaks supersymmetry in a (classically) controlled way.

The background fluxes generically stabilise  complex structure and seven-brane moduli, which can then be assumed to decouple at low energy. In such a case, in \cite{Martucci:2014ska} it was argued that the (implicitly defined) K\"ahler potential describing the moduli sector of the  four-dimensional effective theory has the following simple form
\be\label{Kpot}
K=-3\log a 
\ee   
up to an additional constant which depends on the specific normalisation of the flux superpotential and of the conformal compensator. 

Let us briefly recall the argument leading to (\ref{Kpot}). The metric ansatz (\ref{10dmetric}) is invariant under a simultaneous constant  rescaling of the four-dimensional metric and the compensator:  $\d s^2_4\rightarrow e^{-2\omega}\d s^2_4$, $\Psi\rightarrow e^{\omega}\Psi$. This rigid redundancy is promoted to a local four-dimensional one once the four-dimensional  fields are allowed to have a non-trivial low-energy dynamics. Hence the effective four-dimensional theory is expected to be invariant under arbitrary Weyl transformations. This property, as well as its supersymmetric generalisation \cite{eff1,eff2,Martucci:2014ska}, is accommodated by the superconformal formulation of four-dimensional supergravity,  see for instance \cite{Freedman:2012zz} for a detailed discussion and references to the original papers. Such a formulation has a non-canonical Einstein-Hilbert term 
\be\label{4dsconf}
\frac12 C\int \sqrt{-g_4}\,|\Psi|^2 e^{-\frac13 K} R_4
\ee
where $C$ is some arbitrary positive constant and $K$ is the K\"ahler potential of the ordinary four-dimensional supergravity,  with canonical Einstein-Hilbert term $\frac12 M^2_{\rm P}\int \sqrt{-g_4}\,R_4$, that is obtained by gauge-fixing the superconformal symmetry \cite{Freedman:2012zz}. Clearly $C$ may be reabsorbed in a rescaling of $\Psi$ and can be chosen in order to simplify the relation between the ten- and four-dimensional quantities.   

 By matching (\ref{4dsconf}) with the ten-dimensional  term $\frac{2\pi}{\ell^8_{\rm s}}\int \sqrt{-g_{10}}\, R_{10}$ evaluated on (\ref{10dmetric}), one readily identifies 
\be
C e^{-\frac13 K} =4\pi\int_X e^{-4A}\d\text{vol}_X=4\pi {\rm v}_0  a
\ee
where we have used (\ref{warpsplit}) and (\ref{A0cond}). We can then choose $C=4\pi {\rm v}_0 $ and arrive at (\ref{Kpot}).\footnote{Note that $C=1$ in \cite{Martucci:2014ska} and then the K\"ahler potentials here and there differ by a shift of $3\log(4\pi {\rm v}_0)$.}  With this choice, the gauge-fixing to ordinary supergravity is achieved by setting $\Psi=\frac{M_{\rm P}}{\sqrt{4\pi{\rm v}_0}}e^{\frac{1}{6}K}=\frac{M_{\rm P}}{\sqrt{4\pi{\rm v}_0 a}}$.


\subsection{Hidden  dependence of $G_3$ on the K\"ahler moduli}
\label{sec:modulidep}

In order to make our point as clean as possible, we now 
focus on warped Calabi-Yau compactifications, hence excluding 7-branes while retaining  mobile D3-branes, O3-planes and a non-trivial 
 three-form flux $G_3\neq 0$.  
In this case the axion-dilaton $\tau$ is constant and the internal space can be written as $X=\hat X/\mathbb{Z}_2$, where $\hat X$ is a Calabi-Yau space and  $\mathbb{Z}_2$ represents 
 the orientifold involution. In particular, the different  fields on $X$ uplift to fields on $\hat X$ of  appropriate parity under the orientifold involution.
 For instance, the K\"ahler form $J$ is even  while $G_3$ is odd, and the harmonic forms $\omega_a$ provide a basis of the  integral even cohomology group $H^2_+(X;\mathbb{Z})$. Since  $\tau$ is constant,   $G_3=G^{2,1}+G^{0,3}$ is actually harmonic and then defines an element of the odd cohomology groups $H^{2,1}_-(X)\oplus H^{0,3}_-(X)$. 
Calabi-Yau three-folds have vanishing Hodge numbers  $h^{1,0}=h^{3,2}=0$. Hence the right-hand side of (\ref{primitive}) is always satisfied in cohomology and  does not lead to any constraint  on the K\"ahler moduli. 
 
On the other hand, the condition (\ref{primitive}) introduces in $G_3$ a hidden dependence on the (non-universal) K\"ahler moduli $v^a$ defined in (\ref{Jexp}). Indeed, let us consider an infinitesimal variation $\delta v^a$,  preserving the constraint (\ref{cc}). 
Being the cohomology class of $G_3$ fixed, we can write the corresponding variation as $\delta G_3=\d \delta\calb_2$, with $\delta\calb_2=\delta C_2-\tau\delta B_2$. Since $\delta J=\delta v^a\omega_a$, the preservation of (\ref{primitive}) under K\"ahler moduli deformations requires 
\be\label{prim1}
\delta v^a\omega_a\wedge G_3+ J\wedge \delta G_3=0\ .
\ee
Observing that the possible supersymmetry-breaking $(0,3)$-component $G^{0,3}$ is automatically 
primitive and then is not affected by the variation of the K\"ahler moduli, we can impose $\delta G^{3,0}=\delta G^{1,2}=\delta G^{0,3}=0$, which allow us to restrict $\delta \calb_2$ to be $(1,1)$ and $\delbar$-closed. We can then expand $\delta \calb_2=\delta\calb^{1,1}_{\rm h}+\delbar\delta\Lambda^{1,0}$, where $\delta\calb^{1,1}_{\rm h}$ is harmonic and describes a variation of the $(B_2,C_2)$ moduli, while $\delta\Lambda^{1,0}$ is a complex (1,0)-form. Note that the one-form $\delta\Lambda^{1,0}$ is determined up to a  $\del$-exact piece. We can then expand
\be
\delta\Lambda^{1,0}=\delta v^a \Lambda^{1,0}_a
\ee
with $\Lambda^{1,0}_a$ such that
\be\label{vcontr}
v^a\Lambda^{1,0}_a=0
\ee
up to a $\del$-exact contribution. The condition (\ref{vcontr}) ensures that $\delta v^a\Lambda^{1,0}_a$ is trivial
for  $\delta v^a$ proportional to $v^a$, i.e.\ for deformations orthogonal to the constraint  (\ref{cc}). 
Hence we can set
\be\label{g3var}
\delta G_3=\delta v^a\del\delbar\Lambda^{1,0}_a
\ee
and this in turn  implies that (\ref{prim1}) can be rewritten as 
\be\label{prim15}
J\wedge \del\delbar\Lambda^{1,0}_a=-\omega_a\wedge G_3\ .
\ee
Note that, consistently, both sides of this equation vanish once contracted with $v^a$.  

We may now remove the arbitrary $\del$-exact contribution to $\Lambda_a^{1,0}$  by imposing $\del^\dagger\Lambda^{1,0}_a=0$
 and rewrite (\ref{prim15})  as 
\be\label{prim2}
 \Delta_X\Lambda^{1,0}_a=-2*_X(\omega_a\wedge G_3)
\ee
where $\Delta_X\equiv\d\d^\dagger+\d^\dagger\d\equiv 2(\delbar^\dagger\delbar+\delbar\delbar^\dagger)$ is the usual Laplacian.\footnote{Note that  $\Lambda^{1,0}_a$ can be identified with a corresponding one-form appearing in \cite{Frey:2013bha,Cownden:2016hpf}, see for instance Eq.~(3.39)  in \cite{Cownden:2016hpf}, where the condition (\ref{prim2}) arises as a constraint for a dimensional reduction performed at fixed non-universal K\"ahler moduli. See later for more comments on this point.}
Hodge theorem then implies that $\Lambda^{1,0}_a$ is uniquely determined in terms of $\omega_a\wedge G_3$ (up to a $\del$-exact term, if one removes the gauge-fixing condition $\del^\dagger\Lambda^{1,0}_a=0$). 
From the primitivity condition (\ref{primitive}) it also follows that (\ref{prim2}) actually requires (\ref{vcontr}), showing the mutual consistency of these equations. 

Equation (\ref{g3var}) specifies the way  in which the $G_3$ flux carries the hidden dependence on the K\"ahler moduli. Hence, we can split $G_3$ into the sum  of a fixed moduli-independent  contribution $G^{(0)}_{3}$,  in the same $H^{2,1}(X)\oplus H^{0,3}(X)$ cohomology class of $G_3$, plus a moduli dependent (2,1) $\delbar$-exact piece.  If we locally set $G^{(0)}_{3}=\d\calb^{(0)}_2$, by repeating the same argument followed above
we can locally write $G_{3}=\d\calb_2$ with
 \be\label{Bdec}
 \calb_2=\calb^{(0)}_2+\delbar b^{1,0}+\ell^2_{\rm s}\beta^\alpha\chi_\alpha
 \ee
where $b^{1,0}$ is some globally defined $(1,0)$-form  and $\chi_\alpha$, $\alpha=1,\ldots,h^{1,1}_-$, are odd harmonic $(1,1)$-forms defining a basis of $H^{2}_-(X;\mathbb{Z})$. The coefficients  $\beta^\alpha$ give  the complex parametrisation of the $(B_2,C_2)$-moduli  and  will enter the  four-dimensional effective theory as chiral fields. On the other hand, the one-form $b^{1,0}$ carries the non-trivial dependence of $G_3$ on the non-universal K\"ahler moduli. More explicitly, we can write
\be
G_3=G^{(0)}_{3}+\del\delbar b^{1,0}
\ee
with
\be\label{b1var}
\delta b^{1,0}=\delta v^a \Lambda^{1,0}_a\ .
\ee


\subsection{Chiral coordinates}
\label{sec:chiral}

We are now in the position to revisit the identification of the moduli chiral coordinates  proposed in \cite{Martucci:2014ska}. As we have already observed, the $(B_2,C_2)$ moduli are naturally parametrised by the complex coordinates  $\beta^\alpha$, while the D3-brane moduli are described by a set of coordinates  $Z^i_I$, with $I=1,\ldots, N_{\rm D3}$, which specify the positions of the mobile D3-branes in some local complex coordinate system $z^i$ along $X$.   

On the other hand, the identification of a set of chiral coordinates $\rho_a$, $a=1,\ldots, h^{1,1}_+$, parametrising the universal modulus $a$, the non-universal K\"ahler moduli $v^a$ and the $C_4$-moduli is less obvious. 
In order to detect them, \cite{Martucci:2014ska} used an argument based on supersymmetric probe D3-brane instantons, working under the assumption that $G_3$ is decoupled from the low-energy dynamics. As we have shown in section \ref{sec:modulidep}, such an assumption misses a hidden $G_3$ dependence on the K\"ahler moduli. Let us then revisit the procedure followed in     \cite{Martucci:2014ska}, now taking into account this additional ingredient.

Consider a Euclidean supersymmetric D3-brane, E3-brane for short,  wrapping an effective (even) divisor $D$ and supporting an anti-self-dual world-volume flux $\calf=\frac{1}{2\pi}\ell^2_{\rm s}F_{\rm D3}-B_2|_D$. Its on-shell action is complex and must depend holomorphically on the background chiral moduli. Having the possibility to choose  $h^{1,1}_+$ independent (even) divisors $D_a$, $a=1,\ldots, h^{1,1}_+$, there are enough probe E3-branes for identifying a set of chiral coordinates $\rho_a$, which should enter as an integral linear combination the on-shell  action of any supersymmetric E3-brane. In fact, for our purposes, we can focus on $\Re\rho_a$ and then consider just the real DBI contribution to the E3-brane action. Indeed, the imaginary part of $\rho_a$ depends just on the R-R axionic moduli, which do not appear in (\ref{Kpot}) and must correspond to (perturbative) isometries of the theory. 

Supersymmetry of the E3-brane is equivalent to a generalised calibration condition \cite{eff1,luca1}, which implies that the on-shell bosonic DBI action takes the form
\be\label{E3}
\frac{1}{2\pi}S_{\rm DBI}=\frac12\int_D e^{-4A} J\wedge J-\frac{1}{2\ell_{\rm s}^4}\Im\tau\int_D \calf\wedge \calf\ .
\ee 
We can choose a basis of even divisors $D_a$ which are Poicar\'e dual to the harmonic forms $\omega_a$ and decompose $D=n^a D_a$. Then, we would like to identify $\Re\rho_a$ such that
\be\label{E3exp}
\frac{1}{2\pi}S_{\rm DBI}= n^a\Re\rho_a+(\text{hol}+\overline{\text{hol}})
\ee 
where  $(\text{hol}+\overline{\text{hol}})$  denotes the real part of a holomorphic function of $(Z^i_I,\beta^\alpha)$. 
As in \cite{Martucci:2014ska}, by extracting the dependence of $S_{\rm DBI}$ on the background moduli 
we will  arrive at our definition of $\Re\rho_a$. The new contribution arising from the hidden K\"ahler moduli dependence discussed in section \ref{sec:modulidep} affects only the second term on the right hand-side of (\ref{E3}), so we can focus on that.

By using (\ref{Bdec}) we can decompose $\calf$ as follows
\be
\calf=\calf^{(0)}+\frac{1}{\Im\tau}\big[\Im(\delbar b^{1,0})+\ell^2_{\rm s}\,\Im\beta^\alpha\,\chi_\alpha\big]|_D
\ee
and then, by performing some integrations by parts, we can write
\be
\begin{aligned}
\frac{1}{2\ell_{\rm s}^4}\Im\tau\int_D \calf\wedge \calf=&\,\frac{1}{2\Im\tau}\Im\beta^\alpha\Im\beta^\beta\int_{D}\chi_\alpha\wedge\chi_\beta\\
&+\frac{1}{2\Im\tau\,\ell^4_{\rm s}}\int_D \Big[\Re\big(b^{1,0}\wedge \bar G_3\big)-\frac12\delbar b^{1,0}\wedge\del \bar b^{0,1}\Big]\\
&+\frac{1}{\ell_{\rm s}^2}\Im\beta^{\alpha}\int_D \calf^{(0)}\wedge \chi_\alpha+\frac{1}{2\ell_{\rm s}^4}\Im\tau\int_D \calf^{(0)}\wedge \calf^{(0)}\ .
\end{aligned}
\ee
The last line can be considered as contributing to the $(\text{hol}+\overline{\text{hol}})$ part in  (\ref{E3exp}) and can then be discarded. On the other hand, as shown in \cite{Martucci:2014ska}  and reviewed in appendix \ref{app:chiralm},  the dependence of the first term in (\ref{E3}) on  the background moduli can be made explicit, by writing $\frac12\int_{D_a} e^{-4A} J\wedge J$  as in (\ref{Dadivisor}). 

By combining these results, our guiding condition (\ref{E3exp}) leads to the following definition of the real part of the chiral coordinates $\rho_a$:
\be\label{rho}
\begin{aligned}
\Re\rho_a\equiv&\,\frac12 a\,\cali_{abc}v^b v^c+\frac12\sum_I \kappa_a(Z_I,\bar Z_I;v)+ h_a(v)-\frac{1}{2\Im\tau}\cali_{a\alpha\beta}\Im\beta^\alpha\Im\beta^\beta\\
&-\frac{1}{2\Im\tau\,\ell^4_{\rm s}}\int_{D_a} \Big[\Re\big(b^{1,0}\wedge \bar G_3\big)-\frac12\delbar b^{1,0}\wedge\del \bar b^{0,1}\Big]\ .
\end{aligned}
\ee
In (\ref{rho}) we have introduced the even-odd-odd intersection numbers 
\be
\cali_{a\alpha\beta}\equiv \int_X\omega_a\wedge\chi_\alpha\wedge\chi_\beta
\ee
 and the locally defined potentials $\kappa_a(z,\bar z;v)$  such that
\be
\omega_a=\ii\del\delbar \kappa_a\ .
\ee
These potentials also enter the definition of $h(v)$:
\be\label{defh}
h_a(v)\equiv -\frac{1}{4\pi\ell^4_{\rm s}}\int_X\big(2\pi\kappa_a-\log|\zeta_a|^2\big)\left(\frac{\ii}{2\Im\tau}G_3\wedge\bar G_3 -Q_6^{\rm nd}\right)
\ee
where $\zeta_a$ is the holomorphic section of the line bundle $\calo(D_a)$ such that $D_a=\{\zeta_a=0\}$. By using the formula $\delta^2(D_a)=\frac{\ii}{2\pi}\del\delbar\log|\zeta_a|^2$, this holomorphic section can also be  used to write the last integral in (\ref{rho}) as an integral over the entire internal space $X$:
\be\label{deltause}
\int_{D_a} (\ldots)=\int_X(\ldots)\wedge \delta^{2}(D_a)=\frac{\ii}{2\pi}\int_X(\ldots)\wedge\del\delbar\log|\zeta_a|^2\ .
\ee

The last line in (\ref{rho}), which  does not appear in the corresponding formula of  \cite{Martucci:2014ska}, is the new contribution due to the hidden $G_3$ dependence on the K\"ahler moduli. As a non-trivial consistency condition, the definition (\ref{rho}) should not depend on the choice of the divisor $D_a$ within its equivalence class, up to possible $(\text{hol}+\overline{\text{hol}})$ contributions. This can  be indeed verified by using (\ref{deltause}) and, remarkably, the last line of (\ref{rho}) is actually crucial to prove this property once the $G_3$ dependence on the K\"ahler moduli is taken into account.   

In the following we will need the derivatives of $\Re\rho_a$ with respect to the non-universal K\"ahler moduli $v^a$. These can be computed by  adapting the corresponding calculation  in \cite{Martucci:2014ska}. In particular, a key formula proved  in Appendix A.1 of \cite{Martucci:2014ska} is the following: under a generic variation $\delta v^a$ of the non-universal K\"ahler moduli, we have that
\be\label{kappaderIIB}
\delta \kappa_a(y;v)=2\,\delta v^b \int_{X,y'} G_X(y,y')(J\wedge \omega_a\wedge\omega_b)(y)
\ee 
where $G_X(y,y')=G_X(y',y)$ is the Green's function associated with the internal metric $\d s^6_X$. This Green's function can also be used to express  $e^{-4A_0}$ appearing in (\ref{warpsplit}) as follows:
\be\label{A0green}
e^{-4A_0(y)}=\frac{1}{\ell^4_{\rm s}}\int_{X,y'}G_X(y,y')Q_6(y')\,.
\ee 

By taking (\ref{deltause}), (\ref{g3var}) and (\ref{b1var}) into account, together with (\ref{kappaderIIB}) and (\ref{A0green}), one can verify  that the hidden dependence on the $v^a$'s carried by the $G_3$ flux appearing in (\ref{defh}) and by the terms in the second line of (\ref{rho}) nicely combine, so that we obtain
\be\label{derho}
\delta\Re\rho_a=\calm_{ab}\,\delta v^b
\ee  
with
\be\label{calm}
\calm_{ab}\equiv \int_X e^{-4A}J\wedge \omega_a\wedge\omega_b-\frac{1}{2\Im\tau\ell^4_{\rm s}}\int_X\omega_a\wedge \Re\left(\Lambda^{1,0}_b\wedge \bar G_3\right)\ .
\ee
Notice that the matrix $\calm_{ab}$ does not depend on the arbitrarily chosen divisors $D_a$ and is symmetric. To make the latter property manifest, one may use the fact that it does not depend on the possible $\del$-exact contribution to $\Lambda^{1,0}_b$ either.
One can then fix the gauge $\del^\dagger\Lambda^{1,0}_b=0$ and use  (\ref{prim2}) to write the second integral on the right-hand side of (\ref{calm}) as 
\be\label{posdef}
\int_X\omega_a\wedge \Re\left(\Lambda^{1,0}_b\wedge \bar G_3\right)=\Re\int_X\delbar\Lambda^{1,0}_a\wedge *_X\del\bar\Lambda^{0,1}_b\ .
\ee

Another useful property of $\calm_{ab}$, which can be derived from (\ref{vcontr}) and (\ref{A0cond}),  is the following: 
\be\label{vcalm}
\calm_{ab}v^b=a\,\cali_{abc}v^bv^c=2 {\rm v}_0a(J\lrcorner\omega_a)\ .
\ee


\subsection{Effective Lagrangian}
\label{sec:effective}

Having identified the real part of the chiral fields $\rho_a$, one can proceed with the calculation of the effective Lagrangian following from the K\"ahler potential (\ref{Kpot}).  In particular, the bosonic effective Lagrangian for gravity and the moduli sector is given by
\be\label{lbos}
\call_{\rm bos}=\,\frac12M^2_{\rm P}\,R_4*1-M^2_{\rm P}\,K_{\cali\bar\calj}\, \d\varphi^\cali\wedge *\d\bar\varphi^{\bar\calj}
\ee
with $K_{\cali\bar\calj}\equiv \del_\cali\del_{\bar\calj}K\equiv \frac{\del^2 K}{\del\varphi^\cali\del\bar\varphi^{\bar\calj}}$,
where $\varphi^\cali$ collectively denote the chiral fields $(\rho_a,\beta^\alpha,Z^i_I)$.

In principle, one may invert (\ref{rho}) and write the universal modulus $a$, and then the K\"ahler potential, as a function of
$(\Re\rho_a,\Im\beta^\alpha,Z^i_I,\bar Z^i_I)$. However  in general, unfortunately, it is not possible to find this function and so the general explicit form of the  K\"ahler potential is not known, as it is also the case in the constant warping approximation \cite{Grimm:2004uq}. Nevertheless, as discussed in  appendix \ref{app:IIB}, one can still compute the second derivatives of $K$ appearing in (\ref{lbos}), along the lines of \cite{Grimm:2004uq,Martucci:2014ska}.  The resulting effective bosonic Lagrangian (\ref{lbos}) takes the form 
\be\label{lbos2}
\begin{aligned}
M^{-2}_{\rm P}\call_{\rm bos}=&\,\frac12R_4*1\,-\,\calg^{ab}\nabla\rho_a\wedge *\nabla\bar\rho_b+\frac{1}{4{\rm v}_0a\Im\tau}v^a\cali_{a\alpha\beta}\d\beta^\alpha\wedge *\d\bar\beta^\beta\\
&\,-\frac{1}{2{\rm v}_0 a}\sum_I g_{i\bar\jmath}(Z_I,\bar Z_I)\d Z^i_I\wedge *\d \bar Z^{\bar\jmath}_I \ .
\end{aligned}
\ee
Here the kinetic metric $\calg^{ab}$ is defined as follows
\be
\calg^{ab}\equiv-\frac{1}{4{\rm v}_0a}\left(\calm^{ab}-\frac{1}{2{\rm v}_0 a}v^av^b\right)
\ee 
where $\calm^{ab}$ is the inverse of the matrix $\calm_{ab}$ defined in (\ref{calm}), and we have introduced the covariant exterior derivative
\be
\nabla\rho_a\equiv\d\rho_a-\cala^I_{ai}\d Z^i_I-\frac{\ii}{\Im\tau}\cali_{a\alpha\beta}\Im\beta^\alpha\d\beta^\beta
\ee
 with
\be\label{defcalA}
\cala^I_{ai}\equiv\frac{\del\kappa_a(Z_I,\bar Z_I;v)}{\del Z^i_I}
\ee
denoting the connection along the moduli space of the $I$-th D3-brane. Furthermore  $g_{i\bar\jmath}(z,\bar z)$ is the  metric on the internal space and indeed the last line in  (\ref{lbos2}) matches the kinetic terms obtained by expanding the action of probe D3-branes. 

 Notice also that the inverse of $\calg^{ab}$ is given by 
\be\label{invG}
\begin{aligned}
\calg_{ab}&=-4{\rm v}_0a\,\calm_{ab}+\calm_{ac}\calm_{bd}\,v^cv^d\\
&=4{\rm v}_0a\left[\int_Xe^{-4A}\omega_a\wedge*_X\omega_b+\frac{1}{2\Im\tau\ell^4_{\rm s}}\int_X\omega_a\wedge \Re\left(\Lambda^{1,0}_b\wedge \bar G_3\right)\right]\ .
\end{aligned}
\ee
Equation (\ref{posdef}) shows that the second, manifestly $G_3$ dependent, term in the last line of (\ref{invG}) is positive definite. Hence, such contribution  tends to increase $\calg_{ab}$ or, in other words, to suppress the kinetic metric $\calg^{ab}$.

The effective Lagrangian derived here is almost identical to the Lagrangian derived in \cite{Martucci:2014ska}. The only difference is that the matrix  $\calm_{ab}$  (denoted by $M^{ab}_{\rm w}$ in \cite{Martucci:2014ska})  now contains also the manifestly $G_3$ dependent term appearing in (\ref{calm}). Furthermore,  by looking at the kinetic terms for $\Im\rho_a$ in (\ref{lbos2}), we recognise the terms obtained in \cite{Frey:2013bha,Cownden:2016hpf} by freezing the non-universal K\"ahler moduli and focusing on the $C_4$-moduli, including such explicitly $G_3$ dependent contribution. Hence our effective theory provides the manifestly supersymmetric completion of the results of   \cite{Frey:2013bha,Cownden:2016hpf},  consistently combining the different possible moduli of this class of flux compactifications. Furthermore,   it matches other previous partial results present in the literature, see the discussion in \cite{Martucci:2014ska} and references therein. 

The manifestly supersymmetric structure of our theory allows us to check that it satisfies the no-scale condition \cite{Cremmer:1983bf,Ellis:1983sf} 
\be\label{noscale}
K^{\cali\bar\calj}K_\cali K_{\bar\calj}=3
\ee  
where $K_\cali\equiv\del_{\cali} K$ and $K^{\cali\bar\calj}$ is the inverse of $K_{\bar\cali\calj}$. As in \cite{Martucci:2014ska}, this provides a non-trivial check of the consistency of our theory, since the underlying ten-dimensional vacua are known to allow for a no-scale-like supersymmetry-breaking flux  $G^{0,3}\neq 0$. 

The no-scale condition (\ref{noscale}) can be more easily verified by rewriting it as  $\det A_{\cali\bar\calj}=0$, where $A_{\cali\bar\calj}\equiv\del_\cali\del_{\bar\calj} e^{-\frac13K}=\del_\cali\del_{\bar\calj}  a$. Indeed, by using (\ref{der1}) and (\ref{der2}) it is not difficult to see that, in our case, $\det A_{\cali\bar\calj}\propto \det\left(\frac{\del v^a}{\del \Re\rho_b}\right)=0$. 


\subsection{Formulation with linear multiplets}
\label{sec:linear}

Our effective theory appears somewhat implicit, in the sense that different quantities are only implicitly defined in terms of the chiral fields $\rho_a,\beta^\alpha,Z^i_I$. 
However, its structure becomes more transparent if instead of the chiral multiplets $\rho_a$ we use as elementary fields their
dual linear multiplets $l^a$ -- see for instance \cite{Binetruy:2000zx,Grimm:2005fa} for reviews on the chiral/linear duality. 

Each  linear multiplet  $l^a$ contains a real scalar $l^a$ (denoted by same symbol) and a three-form  field-strength $\calh^a$  as bosonic components. The dualisation to linear multiplets is possible since the K\"ahler potential (\ref{Kpot}) does not depend on the axionic fields $\Im\rho_a$, which can then be dualised to the two-form potentials associated with the field-strengths $\calh^a$. On the other hands the real scalars $l^a$  are given by
\be\label{defl}
l^a=-\frac12\frac{\del K}{\del\Re\rho_a}=\frac{v^a}{2{\rm v}_0a}\ .
\ee
Hence, in the dual picture the universal modulus $a$ and the (constrained) non-universal K\"ahler moduli $v^a$ reorganise themselves into the $h^{1,1}_+$ linear multiplets $l^a$. Indeed (\ref{defl}) can be easily inverted into
\be\label{invl}
a=\frac{1}{2{\rm v}_0}\left(\frac{6{\rm v}_0}{\cali_{abc}\,l^al^bl^c}\right)^{\frac13}\,,\quad v^a=l^a\left(\frac{6{\rm v}_0}{\cali_{bcd}\,l^bl^cl^d}\right)^{\frac13}\ .
\ee

By using (\ref{invl}) one can obtain, through a Legendre transform, the kinetic potential 
\be\label{tildeK0}
\tilde K\equiv\,K+2l^a\Re\rho_a
\ee
which specifies the dual theory. Up to an irrelevant additional constant, this has the following explicit form 
\be\label{tildeK}
\begin{aligned}
\tilde K=&\,\log\Big(\cali_{abc}l^al^bl^c\Big)+\sum_I \hat k(Z_I,\bar Z_I;l)-\frac{1}{\Im\tau}l^a\cali_{a\alpha\beta}\Im\beta^\alpha\Im\beta^\beta\\
&\, -\frac{1}{2\pi\ell^4_{\rm s}}\int_X\big[2\pi \hat k-l^a\log|\zeta_a|^2\big]\left(\frac{\ii}{2\Im\tau}G_3\wedge\bar G_3-Q^{\rm nd}_{6}\right)\\
&-\frac{1}{\Im\tau\,\ell^4_{\rm s}}l^a\int_{D_a} \Big[\Re\big(b^{1,0}\wedge \bar G_3\big)-\frac12\delbar b^{1,0}\wedge\del \bar b^{0,1}\Big]
\end{aligned}
\ee
where 
\be
\hat k(z,\bar z;l)\equiv l^a\kappa_a(z,\bar z;l)
\ee
can be considered as the K\"ahler potential associated with the rescaled K\"ahler form
\be\label{resJ}
\hat J\equiv\ii\del\delbar \hat k\equiv \frac{1}{2{\rm v}_0 a} J= l^a \omega_a
\ee
whose (unconstrained) K\"ahler moduli are given by the bosonic components of the linear multiplets.  

The effective field theory can be derived  from the kinetic potential (\ref{tildeK}), which must be considered as a function of $(l^a,\beta^\alpha,\bar\beta^{\bar\alpha},Z^i_I,\bar Z^{\bar\imath}_I)$.  In particular, the bosonic action is given by \cite{Grimm:2005fa}
\be\label{dualkinetic}
\begin{aligned}
M^{-2}_{\rm P}\call_{\rm linear}=&\,\frac12\,R_4*1+\frac14\,\tilde K_{ab}\left(\d l^a\wedge *\d l^b+\calh^a\wedge *\calh^b\right)-\,\tilde K_{\cali\bar\calj}\d\varphi^{\cali}\wedge *\d\bar\varphi^{\bar\calj}\\
&+\frac\ii2 \,\big(\tilde K_{a\cali} \d\varphi^{\cali}-\tilde K_{a\bar\cali} \d\bar\varphi^{\bar\cali}\big) \wedge \calh^a
\end{aligned}
\ee
where now $\varphi^{\cali}$ collectively denote only the chiral fields $(\beta^\alpha,Z^i_I)$ and we have used the notation for derivatives introduced  in (\ref{noscale}), e.g.\ $\tilde K_{a\cali}\equiv \frac{\del^a\tilde K}{\del l^a\del \varphi^\cali}$, etc. Notice that while computing the derivatives of $\tilde K$ one should again take into account the hidden dependence of $G_3$ on the K\"ahler moduli and hence on the scalars $l^a$. Since the primitivity condition (\ref{primitive}) is unaffected by a possible rescaling of the K\"ahler form, the discussion of section \ref{sec:modulidep} can be easily rephrased by replacing $J\rightarrow \hat J$ and the formulas (\ref{g3var}) and (\ref{b1var}) can be rewritten as
\be
\frac{\del G_3}{\del l^a}=\del\delbar\hat\Lambda^{1,0}_a\,,\quad \frac{\del b^{1,0}}{\del l^a}=\hat\Lambda^{1,0}_a
\ee
respectively, with $\hat\Lambda^{1,0}_a\equiv 2{\rm v}_0a\,\Lambda^{1,0}_a$ such that
\be
\hat\Delta_X\hat\Lambda^{1,0}_a=-2\,\hat*_X(\omega_a\wedge G_3)\ .
\ee
Here and in the following all `hatted' operators are defined in terms of the rescaled K\"ahler metric 
\be\label{resmetric}
\d\hat s^2_X\equiv \frac{1}{2{\rm v}_0 a}\d s^2_X
\ee
associated with the rescaled K\"ahler potential (\ref{resJ}).

The derivatives appearing in (\ref{dualkinetic}) can be explicitly computed and  $\call_{\rm linear}$ assumes the following form
\be
\begin{aligned}
M^{-2}_{\rm P}\call_{\rm linear}=&\,\frac12\,R_4*1-\frac14\,\calg_{ab}\left(\d l^a\wedge *\d l^b+\calh^a\wedge *\calh^b\right)\\
&\,-\,\sum_{I}\hat g_{i\bar\jmath}(Z_I,\bar Z_I)\d Z^i_I\wedge *\d\bar Z^{\bar\jmath}_I+\frac{1}{2\Im\tau}\,l^a\cali_{a\alpha\beta}\d\beta^\alpha\wedge*\d\bar\beta^\beta\\
&\,-\,\Im\left(\cala^I_{ai}\d Z^i_I+\frac{\ii}{\Im\tau}\cali_{a\alpha\beta}\Im\beta^\alpha\d\beta^\beta\right)\wedge\calh^a\ .
\end{aligned}
\ee
In this formula, $\calg_{ab}$ is the same matrix introduced in (\ref{invG}), which can be rewritten as follows
\be
\calg_{ab}=2\int_X e^{-4\hat A}\omega_b\wedge\hat *_X\omega_b+\frac{1}{\Im\tau\ell^4_{\rm s}}\int_X\omega_a\wedge\Re\left(\hat\Lambda^{1,0}_b\wedge \bar G_3\right)
\ee
where the rescaled warp factor
\be
e^{-4\hat A}\equiv (2v_0 a)^2 e^{-4A}=\frac{3}{\cali_{abc}l^al^bl^c}+e^{-4\hat A_0}
\ee
solves the modified equation obtained by substituting  the metric (\ref{resmetric}) in (\ref{warpeq}),   while $e^{-4\hat A_0}$ is the particular solution with vanishing total integral.

Clearly  the use of linear multiplets allows for a more explicitly formulation of the effective theory in terms of the elementary fields, which furthermore have a more direct connection with the geometrical structure of the underlying ten-dimensional vacua. In particular, the rescaled K\"ahler metric (\ref{resmetric}) can be interpreted precisely as the metric `seen' by the mobile D3-branes and naturally enters the other quantities describing the effective theory. In this sense, at least at the perturbative level, the linear multiplet formulation appears more natural than the corresponding formulation in terms of chiral multiplets.



\section{Effective theory of warped M-theory models}
\label{sec:Ftheory}

So far we have assumed the axion-dilaton $\tau$ to be constant, but the results should clearly extend to F-theory models, i.e.\ with non-constant holomorphic $\tau$ and bulk seven-branes. A purely IIB description of such generalisation is complicated by the fact that  $G_3$ transforms non-trivially under SL$(2;\mathbb{Z})$ duality transformations. Furthermore, once seven-branes are introduced, the associated world-volume fluxes are naturally entangled with the $G_3$ fluxes through the Bianchi identities and  must themselves satisfy a primitivity condition. The cohomological structures describing these effects must be appropriately `twisted' in order to take in account the non-trivial SL$(2;\mathbb{Z})$ transformations and are then more complicated with respect to the constant $\tau$ case considered in the previous section. 

On the other hand, these technical difficulties  appear more treatable if addressed from the dual perspective of  M-theory flux compactifications to three dimensions \cite{Becker:1996gj,Becker:2001pm} on elliptically fibered Calabi-Yau four-folds. In such a dual description, the holomorphic axion-dilaton and the seven-branes are geometrised into the non-triviality of the elliptic fibration \cite{Vafa:1996xn} and the IIB three-form and seven-brane fluxes are both represented by the M-theory four-form flux. 

In the following we will show how the same logic followed for type IIB  in the section \ref{sec:preliminaries} and in \cite{Martucci:2014ska} can be easily adapted to the M-theory vacua described in \cite{Becker:1996gj,Becker:2001pm}, by considering  flux compactifications on generic, non-necessarily elliptically fibered, Calabi-Yau four-folds. The effective theory for these kinds of compactifications was first obtained  in the constant warping approximation in \cite{Haack:1999zv,Haack:2001jz}\footnote{See also \cite{Grimm:2010ks} for a discussion relevant for applications to F-theory models.}, while the effect of a weak warping and of higher derivative corrections has been more recently studied  in  \cite{Grimm:2014efa,Grimm:2015mua}. As we will see, our approach allows us to incorporate the effect of the four-form flux, mobile M2-branes and of a possibly strong warping  in a consistent way.  
We will discuss a simple duality check of consistency between the M-theory and IIB results,  leaving a detailed  study of   the implications of our results for F-theory models to the future.

\subsection{Eleven-dimensional structure}
\label{sec:Mstructure}

In the M-theory flux vacua of \cite{Becker:1996gj,Becker:2001pm} the metric takes the form
\be
\d s^2_{11}=\ell^2_{\rm M}|\Phi|^4 e^{4D}\d s^2_3+\ell^2_{\rm M}e^{-2D}\d s^2_Y
\ee 
where $\ell_{\rm M}$ is the M-theory Planck length, $\d s^2_3$ is the flat three-dimensional metric,  $\d s^2_Y$ is the Ricci-flat metric of a Calabi-Yau four-fold $Y$ and the warp factor  $e^{D}$  varies along $Y$. Furthermore 
 $\Phi$ (which has dimension $[\text{mass}]^{\frac12}$) is a constant playing the role of conformal compensator and is fixed by the three-dimensional Einstein frame condition. 
 
 The  M-theory field-strength $F_4$ has the form
 \be
 F_4=\ell^3_{\rm M}|\Phi|^6 \d{\rm vol}_3\wedge \d e^{6D}+G_4
 \ee
where $ \d{\rm vol}_3$ is the volume form associated with $\d s^2_3$.
 The internal $G_4$ flux must be self-dual, $*_YG_4=G_4$.   More precisely, three-dimensional  $\caln=2$  supersymmetry requires the internal $G_4$ flux to be purely $(2,2)$ and primitive,
\be\label{Mprim}
J_{}\wedge G_4=0
\ee
while the possible $(4,0)$ and $(0,4)$ components of $G_4$, also allowed by the self-duality condition, break supersymmetry in a no-scale way. 
In addition,  these vacua can host mobile M2-branes.

Under the above conditions, the equation of motion of  $G_4$ implies that the warp factor is  determined by the Poisson equation
 \be\label{warpeqM}
 \Delta_Y e^{-6D}=\frac{1}{\ell^6_{\rm M}}*_Y Q_8
 \ee
where $Q_8$ is the M2-charge density
\be\label{q8}
Q_8=\frac12 G_4\wedge G_4+\ell^6_{\rm M}\sum_{I\in \text{M2's}}\delta^8_I\,-\ell^6_{\rm M}\,I_8\ .
\ee
Here $I_8$ is a closed 8-form   that is quartic in the curvature and  integrates to $\frac1{24}$ the Euler characteristic of $Y$: $\int_Y I_8=\frac1{24}\chi(Y)$ \cite{Duff:1995wd}. 
As in the  IIB case, once the tadpole condition $\int_X Q_8=0$ is satisfied, the general solution of (\ref{warpeqM}) can be written as
\be\label{Msplit}
e^{-6D}=c+e^{-6D_{0}}
\ee
where $c$ is the universal modulus of these M-theory compactifications -- the counterpart of $a$ in type IIB -- and $e^{-6D_0}$ is the particular solution of (\ref{warpeqM}) that satisfies the condition
\be\label{Mnorm}
\int_Y e^{-6D_{0}}\d\text{vol}_Y=0\ .
\ee
The universal modulus is the breathing mode of the internal space, while the volume of the internal space is not physical and can be fixed to the any given (dimensionless) value
\be\label{Mconst}
\int_Y \d\text{vol}_Y=\frac{1}{4!}\int_Y J_{}\wedge J_{}\wedge J_{}\wedge J_{}\equiv{\rm w}_0\ .
\ee 

\subsection{K\"ahler potential}

The arguments followed in \cite{Martucci:2014ska} and reviewed in section \ref{sec:preliminaries} can be naturally adapted to the warped M-theory compactifications and lead to similar results.
So, we will be sketchy.  First, we assume that the $G_4$-flux stabilises completely the complex structure moduli of $Y$ (i.e.\ the IIB axion-dilaton, complex structure and seven brane moduli in an F-theory context).  One can then invoke the same arguments based on superconformal invariance of \cite{eff1,eff2,Martucci:2014ska}, for instance by using the super-Weyl invariant three-dimensional supergravity described in detail in \cite{Kuzenko:2013uya}. In particular, the three-dimensional Einstein term takes the form
\be\label{Meinstein}
\frac12 C\int  \sqrt{-g_3}\,|\Phi|^2 e^{-\frac14 K} R_3 
\ee
where $K$ denotes the three-dimensional K\"ahler potential and $C$ is an arbitrary positive constant which may be reabsorbed into a rescaling of $\Phi$. 
By matching (\ref{Meinstein}) with the expression obtained by dimensionally reducing the M-theory action leads to the identification
\be\label{MexpK}
 C e^{-\frac14 K}=4\pi\int_Ye^{-6D}\d\text{\rm vol}_Y=4\pi{\rm w}_0\, c
\ee
where, in the second step, we have used the splitting (\ref{Msplit}) and the normalisation condition (\ref{Mnorm}).
Hence, by choosing $C=4\pi{\rm w}_0$, we conclude that the three-dimensional K\"ahler potential takes the following simple form
\be\label{MKpot}
K=-4\log c
\ee
which is completely analogous to (\ref{Kpot}). Fixing the super-Weyl symmetry and imposing a canonical Einstein-Hilbert term $\frac12 M_{\rm P}\int  \sqrt{-g_3}\, R_3 $ require that
\be
\Phi=\left(\frac{M_{\rm P}}{4\pi{\rm w}_0}\right)^{\frac12}e^{\frac18 K}=\left(\frac{M_{\rm P}}{4\pi{\rm w}_0 c}\right)^{\frac12}\ .
\ee 

\subsection{K\"ahler moduli and $G_4$ dependence}
\label{sec:Mkahler}

The Calabi-Yau K\"ahler form $J_{}$ can be expanded as follows
\be
J_{}=u^A \omega_A\ .
\ee
Here $\omega_A$ are harmonic $(1,1)$-forms defining a basis of $H^2(Y;\mathbb{Z})$ and $u^A$ are the non-universal K\"ahler parameters which, because of (\ref{Mconst}), must satisfy the constraint
\be\label{Mcon0}
\frac{1}{4!}\,\cali_{ABCD}\,u^Au^Bu^Cu^D={\rm w}_0
\ee 
where we have introduced the intersection numbers
\be
\cali_{ABCD}\equiv \int_Y \omega_A\wedge \omega_B\wedge\omega_C\wedge\omega_D\ .
\ee
In the following, we will also use local potentials $\kappa_A$ such that
\be\label{Mdefkappa}
\omega_A=\ii\del\delbar\kappa_A\ .
\ee
Geometrically, $e^{-2\pi\kappa_A}$ defines a metric for the line bundle $\calo(S_A)$.   

 The actual (non-massive) K\"ahler moduli  must respect (\ref{Mprim}). This condition can be imposed at a cohomological level by
requiring that 
\be\label{cohoprim}
u^A[\omega_A\wedge G_4]=0 \quad\text{in $H^6(Y)$}\ .
\ee
The $ u^A$'s preserving (\ref{cohoprim}) can be then parametrised in terms of the actual K\"ahler moduli  $v^a$, $a=1,\ldots,N_K$, as follows
\be\label{udec}
u^A=m^A_a\, v^a
\ee
where $\vec m_a=(m^1_a, \ldots, m^{b_2(Y)}_a) $ provide a set of $N_K$ independent vectors such that
\be
m^A_a[\omega_A\wedge G_4]=0\ .
\ee
Since $G_4+\frac12 c_2(Y)$, where $c_2(Y)$ is the second Chern class of $Y$, defines an element of $H^4(Y;\mathbb{Z})$  \cite{Witten:1996md},  the numbers $m^A_a$ can be chosen to be integral: $m^A_a\in\mathbb{Z}$. 

 One can now repeat the arguments of section (\ref{sec:modulidep})
and deduce that a deformation $\delta v^a$ must be accompanied by a deformation of the $G_4$ flux. The corresponding variation $\delta G_4$ can be written as
\be\label{G4var}
\delta G_4=\ii\delta v^a\del\delbar \Lambda^{1,1}_a
\ee
where $\Lambda^{1,1}_a$ are real $(1,1)$-forms, defined up to harmonic and $\del\delbar$-exact pieces.  Each $\Lambda^{1,1}_a$ must satisfy
\be\label{Mort}
v^a\Lambda^{1,1}_a=0
\ee
up to harmonic and $\del\delbar$-exact pieces, and
\be\label{MprimLambda0}
\ii\del\delbar\Lambda^{1,1}_a\wedge J_{}=-\omega_a\wedge G_4
\ee
where
\be
\omega_a\equiv m_a^A\omega_A\ .
\ee

We  can impose that the harmonic component of $\Lambda^{1,1}_a$ vanishes:
\be\label{harmcond}
(\Lambda^{1,1}_{a})_{{\rm harm}}=0
\ee
and   fix the residual degeneracy by requiring that $\d^\dagger \Lambda^{1,1}_a=\del^\dagger\Lambda^{1,1}_a=\delbar^\dagger\Lambda^{1,1}_a=0$. Under these conditions, the primitivity of $G_4$ implies that $\Delta_Y(J\lrcorner\Lambda^{1,1}_a)=0$ and then $\Lambda^{1,1}_a$ is primitive itself:\footnote{Indeed $\Delta_Y(J\lrcorner\Lambda^{1,1}_a)=0$ implies that $J\lrcorner\Lambda^{1,1}_a$ is a constant. On the other hand, by our gauge-fixing conditions, $\Lambda^{1,1}_a$ is a $\d^\dagger$-exact 2-form. Since $[J\lrcorner,\d^\dagger]=0$, also $J\lrcorner\Lambda^{1,1}_a$ is $\d^\dagger$-exact and then it cannot be a non-vanishing constant.}
\be
J\lrcorner\Lambda^{1,1}_a=0\ .
\ee   
In turn this property can be used to rewrite  (\ref{MprimLambda0}) in the form
\be\label{MprimLambda}
\Delta_Y \Lambda^{1,1}_a =-2 *_Y(\omega_a\wedge G_4)\ .
\ee
Note that  (\ref{Mort})  is not an independent condition but, because of the primitivity of $G_4$, it is actually required by  (\ref{MprimLambda}).

 Along the same lines, we can split 
\be
G_4=G^{(0)}_4+\ii \del\delbar \calc^{1,1}
\ee
where $G^{(0)}_4$ is any fixed element of $H^{4,0}(Y)\oplus H^{2,2}(Y)\oplus H^{0,4}(Y)$ in the same class of $G_4$ while  $\calc^{1,1}$ is  a globally defined real  (1,1)-form  encoding the  K\"ahler moduli dependence:
\be\label{C2var}
\delta\calc^{1,1}=\delta v^a\Lambda^{1,1}_a\ .
\ee
We can then locally write $G^{(0)}_4=\d C^{(0)}_3$ and $G_4=\d C_3$ with
\be\label{C3dec}
C_3=C^{(0)}_3-\frac{\ii}{2}\del \calc^{1,1}+\frac{\ii}{2}\delbar \calc^{1,1}+\ell^3_{\rm M}(\beta^\alpha\lambda_{\alpha}+\bar\beta^{\bar\alpha}\bar\lambda_{\bar\alpha})
\ee 
where the $\lambda_{\alpha}$, $\alpha=1,\ldots, b_3(Y)$, define a harmonic basis of $H^{1,2}(Y)$. 

\subsection{Chiral coordinates}

We now need to describe the moduli space in chiral coordinates. The M2-brane moduli can be parametrised by their positions $Z^i_I$ in some local complex coordinates $z^i$ along $Y$. The moduli of the $C_3$  gauge potential  can be identified with the  $\beta^\alpha$ appearing in (\ref{C3dec}). On the other hand, the precise form of the chiral coordinates $\rho_a$ of the K\"ahler and $C_6$ moduli (where $C_6$ is the potential of $F_7=*_{11}F_4$) is less obvious, as for the K\"ahler/$C_4$ moduli in type IIB. 

One can proceed as in  \cite{Martucci:2014ska} and in section \ref{sec:chiral}. The relevant instantons are given by Euclidean M5-branes. As in type IIB, the perturbative effective Lagrangian does not depend on the $C_6$ axions, and hence on $\Im\rho_a$. We can then focus on $\Re\rho_a$, which must be detected by the real part of the action of a probe supersymmetric M5-brane wrapping an effective divisor $S\subset Y$.
This contains a warped volume contribution proportional to  
\be\label{firstM5}
\frac1{3!}\int_S e^{-6D}J_{}\wedge J_{}\wedge J_{}
\ee
analogous to the first term on the right-hand side of (\ref{E3}). Hence,  $\Re\rho_a$ must contain a contribution similar to (\ref{firstM5}) for an appropriately chosen divisor $S$. Indeed, this parametrisation has been used in \cite{Grimm:2015mua} and shown to be consistent with a direct dimensional reduction. 

More precisely, the contribution of (\ref{firstM5}) to $\Re\rho_a$ can be obtained by choosing the integration divisor 
\be
S_a=m_a^A S_A
\ee
 where $S_A$, $A=1,\ldots, b_2(Y)$, are a set  of divisors which are Poincar\'e dual to the integral harmonic forms $\omega_A$. As shown in appendix \ref{app:Mrho}, by discarding some 
(hol+$\overline{\text{hol}})$ contribution one is then led to identify part of $\Re\rho_a$ with
\be\label{Mfirstcont}
\frac1{3!} c\,\cali_{aABC}u^A u^B u^C+\frac12\sum_I \kappa_a(Z_I,\bar Z_I;v)+ h_a(v)
\ee
where $\cali_{aABC}\equiv m_a^D\cali_{ABCD}$,  $\kappa_a(z,\bar z;v)\equiv m_a^A\kappa_A(z,\bar z;v)$ [see equation (\ref{Mdefkappa})] 
and $h_a(v)$ is defined as follows
\be\label{Mhdef}
h_a(v)\equiv\frac1{4\pi\ell^6_{\rm M}}\int_Y (2\pi\kappa_a-\log|\zeta_a|^2)\left[\frac1{2}G_4\wedge G_4-\ell^6_{\rm M}I_8\right]
\ee
where $\zeta_a(z)$ is a holomorphic section of $\calo(S_a)$ which vanishes on $S_a$. 

The M5-brane action contains other terms in addition to (\ref{firstM5}). First, there may be higher order derivative corrections,  similar to the  $I_8$ appearing in (\ref{Mhdef}), which may carry a dependence on the moduli, in particular on the K\"ahler ones. Such higher order corrections are not the main focus of the present paper and so we will neglect them, while considering  $I_8$ as a non-dynamical contribution to the M2-charge density. One may incorporate their effect along the lines of \cite{Grimm:2014efa,Grimm:2015mua} which showed how corrections of this kind are in fact necessary in order to accommodate  M-theory higher derivative contributions to the effective theory.  

Rather, we concentrate on the   $G_4$ dependent  contribution to the M5-brane action, which is present already at the lowest derivative level. This term should arise from the contribution of the world-volume 3-form flux $T_3=\d A_2+C_3$  supported on the M5-brane which is, however,  self-dual and then does not admit a simple Lagrangian description \cite{Witten:1996hc,Pasti:1997gx,Bandos:1997ui}. One can then follow a simpler strategy, which  identifies such a term by requiring that, as in the type IIB case,  the definition of $\Re\rho_a$ depends  on the choice of the divisors $S_A$ (within their linear equivalence classes) at most by a (hol+$\overline{\rm hol}$) term and matches the IIB result under duality. These conditions lead to the definition
\be\label{Mrho}
\begin{aligned}
\Re\rho_a\equiv&\frac1{3!} c\,\cali_{aABC}u^A u^B u^C+\frac12\sum_I \kappa_a(Z_I,\bar Z_I;v)+ h_a(v)\\
&+\,\frac12\calt_{a\alpha\bar\beta}\,\beta^\alpha\bar\beta^{\bar\beta}+\frac{1}{2\ell^6_{\rm M}}\int_{S_a}\left(\calc^{1,1}\wedge G_4+\frac\ii2 \del\calc^{1,1}\wedge\delbar\calc^{1,1}\right)
\end{aligned}
\ee
where
\be
\calt_{a\alpha\bar\beta}\equiv\ii \int_Y\omega_a\wedge\lambda_\alpha\wedge\bar\lambda_{\bar\beta}\ .
\ee
 See subsection \ref{sec:matching} below for a discussion on the matching with type IIB side.

Relegating the details of the calculation to appendix \ref{app:delrho},
we obtain that, under a deformation of the non-universal K\"ahler moduli $\delta v^a$ [preserving (\ref{Mcon0}) and (\ref{cohoprim})], $\Re\rho_a$ transforms as follows
\be\label{Mderho}
\delta\Re\rho_a=\caln_{ab}\,\delta v^b
\ee
where
\be\label{calnab}
\caln_{ab}\equiv \frac12\int_Ye^{-6D}J\wedge J\wedge \omega_a\wedge \omega_b+\frac{1}{2\ell^6_{\rm M}}\int_Y \omega_a\wedge\Lambda^{1,1}_b\wedge G_4 \ .
\ee
Note that, in $\delta\Re\rho_a$, the contributions due to the  $G_4$ dependence on the K\"ahler moduli coming from $h_a(v)$ and from the second line of (\ref{Mrho}) nicely combine, so that the final result does not depend on the arbitrarily chosen divisors $S_a$.

We observe that equation (\ref{MprimLambda}) can be used to deduce that $\caln_{ab}$ contains the 
manifestly symmetric and positive definite  contribution
\be\label{Mrew}
-\frac{1}{2\ell^6_{\rm M}}\int_Y\Lambda^{1,1}_a\wedge \omega_b\wedge G_4=\frac1{4\ell^6_{\rm M}}\int_Y\d\Lambda^{1,1}_a\wedge *_Y\d\Lambda^{1,1}_b\ .
\ee
Note also that by contracting (\ref{calnab}) with $v^a$ and using the primitivity of $G_4$ one gets the useful identity  
\be\label{Ncontr}
\caln_{ab}\,v^b=\frac12\,c\,\cali_{a ABC}u^Au^Bu^C=3{\rm w}_0c\, (J\lrcorner \omega_a)\ .
\ee

\subsection{Effective Lagrangian}

By using the above results, one can compute the effective Lagrangian associated  with the K\"ahler potential (\ref{MKpot}),  as we did  in section \ref{sec:effective} for the type IIB case. The formula (\ref{lbos}) is valid also for  three-dimensional  $\caln=2$ theories, up to substituting $M^2_{\rm P}$ with $M_{\rm P}$, see for instance \cite{Kuzenko:2013uya}. Hence, by using the results of appendix \ref{app:Meff}, one obtains  the following Lagrangian for the  moduli sector of the effective theory:   
\be\label{Mlbos2}
\begin{aligned}
M^{-1}_{\rm P}\call_{\rm M}=&\frac12R_4*1\,-\,\calg^{ab}\nabla\rho_a\wedge *\nabla\bar\rho_b-\frac{1}{2{\rm w}_0c}v^a\calt_{a\alpha\bar\beta}\d\beta^\alpha\wedge *\d\bar\beta^{\bar\beta}\\
&\,-\frac{1}{2{\rm w}_0 c}\sum_I g_{i\bar\jmath}(Z_I,\bar Z_I)\d Z^i_I\wedge *\d \bar Z^{\bar\jmath}_I 
\end{aligned}
\ee
We have introduced the covariant derivative 
\be
\nabla\rho_a\equiv\d\rho_a-\cala^I_{ai}\d Z^i_I-\calt_{a\alpha\bar\beta}\,\bar\beta^{\bar\beta}\d\beta^\alpha
\ee
where 
\be\label{Mcala}
\cala^I_{ai}\equiv\frac{\del\kappa_a(Z_I,\bar Z_I;v)}{\del Z^i_I}\ 
\ee
is the connection  along the moduli space of the $I$-th M2-brane. The kinetic metric $\calg^{ab}$ is defined as follows
\be
\calg^{ab}\equiv-\frac{1}{4{\rm w}_0c}\left(\caln^{ab}-\frac{1}{3{\rm w}_0 c}v^av^b\right)
\ee
where $\caln^{ab}$ is the inverse of the matrix $\caln_{ab}$ defined in (\ref{calnab}). Furthermore, note that the contribution appearing in the last line of (\ref{Mlbos2}) contains the Calabi-Yau metric $g_{i\bar\jmath}$ and perfectly matches the kinetic terms obtained by considering probe M2-branes  on the M-theory vacua described  in  section (\ref{sec:Mstructure}).

The kinetic metric $\calg^{ab}$ is the inverse of
\be\label{MinvG}
\begin{aligned}
\calg_{ab}&=-4{\rm w}_0c\,\caln_{ab}+\frac49\caln_{ac}\,\caln_{bd}\,v^cv^d\\
&= 4{\rm w}_0 c\left(\int_Y e^{-6D}\omega_a\wedge *_Y\omega_b-\frac{1}{2\ell^6_{\rm M}}\int_Y \Lambda^{1,1}_a\wedge \omega_b\wedge G_4\right)\ .
\end{aligned}
\ee
The first warped term in the second line was also obtained  in \cite{Grimm:2014efa,Grimm:2015mua} by a direct  dimensional reduction in a weak warping regime. However, as it happens in the  IIB case, we see that  it is valid also for a possible strong warping and that it must be completed by a $G_4$ dependent contribution. Note that such a contribution can be written in a manifestly symmetric and positive definite form by using (\ref{Mrew}).  Hence, as in type IIB, this contribution of the flux tends to lower the value of the kinetic matrix $\calg^{ab}$. 

This supersymmetric theory satisfies the three-dimensional no-scale condition\footnote{The no-scale property can be more easily checked by verifying that the  matrix $A_{\cali\bar\calj}\equiv \del_\cali\del_{\bar\calj}e^{-\frac14 K}\equiv \del_\cali\del_{\bar\calj}c$ is degenerate.}
\be\label{Mnoscale}
K^{\cali\bar\calj}K_{\cali}K_{\bar\calj}=4
\ee
where $K^{\cali\bar\calj}$ is the inverse of $K_{\bar\cali\calj}$ and we are adopting the usual notation: $K_{\cali}\equiv \del_\cali K\equiv \frac{\del K}{\del\varphi^\cali}$, $K_{\cali\bar\calj}\equiv \del_\cali \del_{\bar\calj}K$ etc.,  with $\varphi^\cali=(\rho_a,Z^i_I,\beta^\alpha)$.  This is indeed consistent with the possibility of having supersymmetry breaking components $G^{4,0}+G^{0,4}\neq 0$, which would  induce a non-vanishing contribution to the superpotential of the three-dimensional theory \cite{Gukov:1999ya}. The condition (\ref{Mnoscale}) ensures the vanishing of  the vacuum expectation value of the three-dimensional potential -- see for instance \cite{Kuzenko:2013uya} for the component form of the action.

Finally, one may highlight the geometrical interpretation of the effective theory by  dualising the chiral multiplets $\rho_a$ into vector multiplets. We do not present the details of such a dualisation, which would be the three-dimensional analog   of what was presented in section \ref{sec:linear} for type IIB compactifications to four dimensions. A more explicit discussion on this duality in our same context can be found  in \cite{Grimm:2015mua}.


\subsection{Matching with type IIB}
\label{sec:matching}

By taking elliptically fibered Calabi-Yau fourfolds and $G_4$-fluxes satisfying appropriate transversality conditions, the M-theory flux compactifications discussed in the present section are dual to type IIB F-theory compactifications to four dimensions. Leaving  to the future a detailed discussion on  the implications  of our results  in the context of F-theory compactifications, we now give  just  a simple check
of the compatibility between the M-theory results of the present section and the IIB results of section \ref{sec:preliminaries}. 

Indeed, one can locally assume an approximately  factorised structure $Y\simeq X\times T^2_\tau$, where $T^2_\tau$ is a two-dimensional torus with approximately constant complex structure $\tau$. In such a case the duality relations between the M-theory and IIB (local) quantities are easily formulated, see for instance \cite{Denef:2008wq}. In particular, we are interested in the duality relation between the terms involving the $G_3/G_4$ fluxes in the definitions of $\rho_a$ in type IIB/M-theory  respectively.

Writing the $T^2_\tau$ metric as
\be
\d s^2_{T^2_\tau}=\frac{L^2}{\Im\tau}\d z\d\bar z
\ee
where $z=x+\tau y$, with periodicities $x\simeq x+1$ and $y\simeq y+1$, the above factorised structure corresponds to taking $\d s^2_Y\simeq \d s^2_X+\d s^2_{T^2_\tau}$. $L^2$ represents the volume of $T^2_\tau$ 
which is eventually sent to zero in the F-theory limit. Then, as in \cite{Denef:2008wq},  the M-theory/IIB  $G_4/G_3$ fluxes are related as follows 
\be\label{G4G3}
G_4=\frac{L}{\Im\tau}\Im\left(\d\bar z\wedge G_3\right)\ .
\ee 
By consistency, the relation between $b^{1,0}$ and  $\calc^{1,1}$  appearing in  (\ref{Bdec}) and (\ref{C3dec}), respectively, is\footnote{The complete relation between the potentials (\ref{C3dec}) and (\ref{Bdec}) is $C_3=-\frac{L}{\Im\tau}\Im\left(\d\bar z\wedge \calb'_2\right)$, with $\calb'_2=\calb_2-\frac12\d b^{1,0}$, $C^{(0)}_3=-\frac{L}{\Im\tau}\Im\big(\d\bar z\wedge \calb^{(0)}_2\big)$ and $\lambda_\alpha=\frac{\ii}{2\Im\tau}\d\bar z\wedge \chi_\alpha$. Furthermore, if $\omega_a$ is Poincar\'e dual to a vertical divisor, then $\calt_{a\alpha\beta}=-\frac1{2\Im\tau}\cali_{a\alpha\beta}$.} 
\be\label{C2b1}
\calc^{1,1}=-\frac{L}{\Im\tau}\Re\left(\d\bar z\wedge b^{1,0}\right)\ .
\ee

Let us now choose as M-theory divisors $S_a$ the vertical divisors in $Y$ obtained by `attaching' the $T^2_\tau$ fibre over the divisors $D_a$ in the base $X$. Then, by plugging  (\ref{G4G3}) and (\ref{C2b1}) into  the last term appearing in the definition (\ref{Mrho}) of $\Re\rho_a$ and performing the integration over the $T^2_\tau$ fibre in $S_a$, it is easy to check that
\be\label{last}
\begin{aligned}
&\frac{1}{2\ell^6_{\rm M}}\int_{S_a}\left(\calc^{1,1}\wedge G_4+\frac\ii2 \del\calc^{1,1}\wedge\delbar\calc^{1,1}\right)\\
&=-\frac{L^2}{2\ell^6_{\rm M}\Im\tau}\int_{D_a}\Big[\Re\big(b^{1,0}\wedge \bar G_3\big)-\frac12\delbar b^{1,0}\wedge\del \bar b^{0,1}\Big]\ .
\end{aligned}
\ee
Upon using the  identification $\ell^3_{\rm M}=L\ell^2_{\rm s}$ \cite{Denef:2008wq},
we see that (\ref{last}) exactly reproduces the last term in the definition (\ref{rho}) of  $\Re\rho_a$ in type IIB.
This shows how such contributions, which are directly related to the manifestly flux-dependent contributions to the respective effective Lagrangians,  are indeed perfectly compatible under duality. Similar consistency  checks   for the  other terms appearing in (\ref{rho}) and (\ref{Mrho}) can also be  performed.

\bigskip

\section{Conclusions}

In this paper we have studied  the effective theory of type IIB/M-theory warped  flux compactifications, completing and extending to the M-theory case the results of \cite{Martucci:2014ska}. The type IIB case is discussed in section \ref{sec:preliminaries} while the M-theory case is considered in section \ref{sec:Ftheory}. The results are very similar and can be indeed  related by duality. 

In our discussion, a  key role is played by the hidden dependence of the $G_3/G_4$ fluxes on the K\"ahler moduli.
Such a dependence does not directly affect the simple formulas (\ref{Kpot})/(\ref{MKpot}) for the K\"ahler potential. Rather,
 it can be taken into account just by including an additional $G_3/G_4$ dependent term -- given by the last line of equations (\ref{rho})/(\ref{Mrho}) -- in the definition of the chiral coordinates $\rho_a$ parametrising the K\"ahler moduli. This  produces a manifestly $G_3/G_4$ dependent  contribution to the effective Lagrangian (\ref{lbos2})/(\ref{Mlbos2}). More precisely, this contribution appears in the second term of the inverse kinetic matrix (\ref{invG})/(\ref{MinvG}) and is always positive, hence having a `suppression effect' on the kinetic matrix $\calg^{ab}$ of the chiral fields $\rho_a$. Our results  reproduce, complete and supersymmetrise  the bosonic effective action obtained in \cite{Frey:2013bha,Cownden:2016hpf} by dimensional reduction of IIB compactifications at fixed non-universal  K\"ahler moduli.

There are several aspects that remain to be explored, including the following ones:
\begin{itemize}

\item Warping and fluxes apparently `break' the cohomological nature of the  effective theories obtained in the constant warping approximation. In particular, for the moment,  the evaluation of the explicit form of the effective theory seems to require a case by case study.  It would be very useful to develop efficient,  model-independent topological/cohomological/algebraic-geometrical techniques to compute  the explicit  structure of our effective theories. This would allow for a better understanding  of the physical and phenomenological implications of our results, both at a qualitative and a quantitative level.    

\item It would be important to further study the applications of our results to the  phenomenologically more relevant framework of F-theory flux compactifications, by combining both the type IIB and M-theory perspectives, as for instance done in  \cite{Denef:2008wq,Grimm:2010ks} in the constant warping approximation. The present paper and \cite{Martucci:2014ska} assume dynamically frozen axion-dilaton and complex structure and focus on the remaining moduli sector, but it would be clearly worthwhile  to  incorporate gauge  and charged matter sectors into the effective theory, as well as possible dynamical complex structure  moduli which may be present in the low-energy spectrum.

\item  In deriving our effective theories we have used the standard two-derivative type IIB/M-theory supergravity. It would be interesting to understand how to incorporate higher-derivative corrections, for instance combining our results with  those of the papers \cite{Grimm:2014efa,Grimm:2015mua}, in which this problem has been investigated for weakly warped M-theory compactifications. 
Higher derivative corrections can play an important role in phenomenological models -- see for instance the reviews  \cite{Denef:2008wq,Maharana:2012tu} -- and so it would be desirable to better understand how to combine them  with the effects of  warping, fluxes and branes.

\item Even though our effective theories have a natural applicability to string phenomenology, they may  also be useful in the somewhat different context of the AdS/CFT correspondence. Indeed, as discussed in \cite{Martucci:2016hbu}, by taking a rigid/decompactification limit thereof, one can obtain the `holographic effective field theory' of strongly coupled quantum field theories. Ref.~\cite{Martucci:2016hbu} considered type IIB holographic models in which the non-compact internal space is an asymptotically conical Calabi-Yau three-fold and the warping is only due  to mobile D3-branes, but the same logic can be applied to more general type IIB/M-theory holographic models for which the results of the present paper can be relevant.   


\end{itemize}


\vspace{1cm}

\centerline{\large\em Acknowledgments}

\vspace{0.4cm}

\noindent I would like to thank Andrew Frey for correspondence and Dmitri Sorokin for  comments on the manuscript. This work was partially supported by the Padua University
Project CPDA144437.

\vspace{0.5cm}


\begin{appendix}

\section{IIB-theory effective theory: some details}
\label{app:IIB0}

In this appendix we collect some technical details regarding the derivation of the effective theory of the IIB compactifications considered in section \ref{sec:preliminaries}.  

\subsection{Explicit form of the warped divisor volume}
\label{app:chiralm}

  In \cite{Martucci:2014ska} it was shown how,  given a divisor $D\simeq n^a D_a$, one can express the integral
\be\label{wD}
\frac12\int_D e^{-4A}J\wedge J
\ee
in terms of the background moduli. Let us briefly review this result and its derivation. 

First one can use (\ref{warpsplit}) and (\ref{Jexp}) to write (\ref{wD}) as follows
\be\label{intDdec}
\frac12a\, n^a\cali_{abc}v^bv^c+\frac12\int_X e^{-4A_0}J\wedge J\wedge \delta^2(D)
\ee
where $\delta^2(D)$ is a delta-like 2-form localised on the divisor $D$. This can be written as $\delta^2(D)=\frac{\ii}{2\pi}\del\delbar\log|\zeta_D|^2$, with  $\zeta_D$ being  a section of the line bundle $\calo(D)$ that vanishes on $D$, so that
\be
J\lrcorner \delta^2(D)=-\frac{1}{4\pi}\Delta\log|\zeta_D|^2\,. 
\ee
Furthermore, since $\omega_a=\ii\del\delbar\kappa_a$ is harmonic, the scalar
\be
d_a\equiv J\lrcorner \omega_a=-\frac12\Delta\kappa_a 
\ee  
is harmonic and hence constant along $X$. We can then use the identity 
\be
J\lrcorner \delta^2(D)=\frac{1}{4\pi}\Delta\left(2\pi n^a \kappa_a-\log|\zeta_D|^2\right)+n^ad_a\,. 
\ee
and (\ref{A0cond}) to rewrite the second term of (\ref{intDdec}) as
\be\label{iii}
\int_X e^{-4A_0}[J\lrcorner \delta^2(D)]\d{\rm vol}_X=\frac1{4\pi}\int_X e^{-4A_0}\Delta\left(2\pi n^a\kappa_a-\log|\zeta_D|^2\right)\d{\rm vol}_X\,.
\ee
Now, the key point is that the local potential $n^a\kappa_a$ defines a metric $e^{-2\pi n^a\kappa_a}$ on the line bundle  $\calo(D)$ and then the combination $2\pi n^a\kappa_a-\log|\zeta_D|^2$ is globally defined. This implies that we can integrate the right-hand side of (\ref{iii}) by parts and use (\ref{warpeq})  to get  the identity
\be
\int_X e^{-4A_0}[J\lrcorner \delta^2(D)]\d{\rm vol}_X=\frac{1}{\ell^4_{\rm s}}\int_D\left(2\pi n^a\kappa_a-\log|\zeta_D|^2\right)Q_6\ .
\ee
By using (\ref{D3charge}), we then conclude that (\ref{wD}) with $D\equiv D_a$ (hence $n^b=\delta^b_a$)   can be rewritten as
\be\label{Dadivisor}
\frac12a\,\cali_{abc}v^bv^c+\frac12\sum_I\kappa_a(Z_I,\bar Z_I; v)+h_a(v)-\sum_I[\log\zeta_a(Z_I)+\log\bar \zeta_a(\bar Z_I)]
\ee
where $h_a(v)$ is defined in (\ref{defh}). The last term in (\ref{Dadivisor}) is $(\text{hol}+\overline{\text{hol}})$ and then, by using this formula in (\ref{E3}) and (\ref{E3exp}), we can discard it and identify the first three terms appearing on the right-hand side of (\ref{rho}). 

\subsection{Derivation of kinetic terms for IIB compactifications}
\label{app:IIB}

In this appendix we discuss the derivation of the effective Lagrangian (\ref{lbos2})
from (\ref{Kpot}) and (\ref{rho}).

We  start from some useful formulas for the implicit derivatives of the moduli. 
Consider the tautological identities $\frac{\del\Re\rho_a}{\del\Re\rho_b}=\delta^b_a$,  $\frac{\del\Re\rho_a}{\del Z^i_I}=0$ and  $\frac{\del\Re\rho_a}{\del \beta^\alpha}=0$. By using (\ref{derho}) we can rewrite them as follows
\begin{subequations}\label{inv}
\begin{align}
\frac12\cali_{acd}v^cv^d\frac{\del a}{\del\Re\rho_b}+\calm_{ac}\frac{\del v^c}{\del\Re\rho_b}&=\delta^b_a\label{inv1}\\
\frac12\cali_{abc}v^bv^c\frac{\del a}{\del Z^i_I}+\calm_{ab}\frac{\del v^b}{\del Z^i_I}+\frac12\cala^I_{ai}&=0\label{inv2}\\
\frac12\cali_{abc}v^bv^c\frac{\del a}{\del \beta^\alpha}+\calm_{ab}\frac{\del v^b}{\del \beta^\alpha}+\frac{\ii}{2\Im\tau}\cali_{a\alpha\beta}\Im\beta^\beta&=0\label{inv3}
\end{align}
\end{subequations}
where $\cala^I_{ai}$ is defined in (\ref{defcalA}). By contracting (\ref{inv}) with $v^a$ and using  (\ref{vcalm}) together with (\ref{cc}) and its corollary  $\cali_{abc}v^av^b\frac{\del v^c}{\del\Re\rho^d}=0$,  we get
\be\label{der1}
\frac{\del a}{\del \Re\rho_a}=\frac{1}{3{\rm v}_0}v^a\,,\quad \frac{\del a}{\del Z^i_I}=-\frac{1}{6{\rm v}_0}v^a\cala^I_{ai}\,,\quad \frac{\del a}{\del \beta^\alpha}=-\frac{\ii}{6{\rm v_0}\Im\tau}v^a\cali_{a\alpha\beta}\Im\beta^\beta\ .
\ee
Using these relations back in (\ref{inv}) and taking again (\ref{vcalm}) into account, one also gets
\be\label{der2}
\begin{aligned}
\frac{\del v^a}{\del\Re\rho_b}&=\calm^{ab}-\frac{1}{6{\rm v}_0 a}v^a v^b\\
\frac{\del v^a}{\del Z^i_I}&=-\frac{1}{2}\left(\calm^{ab}-\frac{1}{6{\rm v}_0 a}v^a v^b\right)\cala^I_{bi}\\
\frac{\del v^a}{\del \beta^\alpha}&=-\frac{\ii}{2\Im\tau}\left(\calm^{ab}-\frac{1}{6{\rm v}_0 a}v^a v^b\right)\cali_{b\alpha\beta}\Im\beta^\beta
\end{aligned}
\ee
where $\calm^{ab}$ is the inverse of $\calm_{ab}$.

Consider now the K\"ahler potential (\ref{Kpot}). By taking into account that it does not depend on $\Im\rho_a$ and $\Re\beta^\alpha$ and  and using (\ref{der1}) we can compute
\be
\frac{\del K}{\del\rho_a}=-\frac{v^a}{2{\rm v}_0a}\,, \quad \frac{\del K}{\del Z^i_I}=\frac{1}{2{\rm v}_0 a}v^a\cala^I_{ai}\,, \quad \frac{\del K}{\del \beta^\alpha}=\frac{\ii}{2 {\rm v}_0a\Im\tau}v^a\cali_{a\alpha\beta}\Im\beta^\beta\ .
\ee
Then, by using again (\ref{der1}) together with (\ref{der2}) and (\ref{kappaderIIB}), we obtain the second derivatives of $K$:
\be
\begin{aligned}
\frac{\del^2 K}{\del\rho_a\del\bar\rho_b}&=-\frac{1}{4{\rm v}_0 a}\left(\calm^{ab}-\frac{1}{2{\rm v}_0 a}v^av^b\right)\equiv \calg^{ab}\,,\\
\frac{\del^2 K}{\del\rho_a\del\bar Z^{\bar\imath}_I}&=  -\calg^{ab}\bar\cala^I_{b\bar\imath}\,, \quad~~~~~~
\frac{\del^2 K}{\del Z^i_I\del\bar Z^{\bar\jmath}_J}= \calg^{ab}\cala^I_{a i}\bar\cala^J_{b\bar\jmath}+\frac{1}{2{\rm v}_0 a}g_{i\bar\jmath}(Z_I,\bar Z_I)\,,  \\
\frac{\del^2 K}{\del\rho_a\del\bar\beta^\alpha}&=  \frac{\ii}{\Im \tau}\calg^{ab}\cali_{b\alpha\beta}\Im\beta^\beta\,, \quad~~~~~~
 \frac{\del^2 K}{\del\beta^\alpha\del\bar Z^{\bar\imath}_I}=\frac{\ii}{\Im\tau}\calg^{ab}\bar\cala^I_{a\bar\imath}\cali_{b\alpha\beta}\Im\beta^\beta\,,\\
\frac{\del^2 K}{\del\beta^\alpha\del\bar\beta^\beta}&=-\frac{1}{4{\rm v}_0a\Im\tau}v^a\cali_{a\alpha\beta}+\frac{1}{(\Im\tau)^2}\calg^{ab}\cali_{a\alpha\gamma}\cali_{b\beta\delta}\Im\beta^\gamma\Im\beta^\delta\,.
\end{aligned}
\ee
These formulas can be used to compute the effective theory. In particular, the bosonic effective Lagrangian takes the  form (\ref{lbos2}).

\section{M-theory effective theory: some details}

In this appendix we derive some formulas presented in section \ref{sec:Ftheory}.

\subsection{Warped divisor volumes in M-theory}
\label{app:Mrho}

We want to make more explicit the dependence of (\ref{firstM5}) on the background moduli, following \cite{Martucci:2014ska}. 
First we can use  (\ref{Msplit}) and the homological decompositions $[S]=m^A [D_A]$  to rewrite (\ref{firstM5}) as follows:
\be\label{Mwarpdiv}
\frac1{3!}\int_Se^{-6D}J\wedge J\wedge J=\frac1{3!} c\,m^A\cali_{ABCD}v^Bv^Cv^D+\int_Ye^{-6D_0}[J\lrcorner\delta^2(S)]\d\text{vol}_Y\ .
\ee
One can now use the identity $\delta^2(S)=\frac{1}{2\pi}\ii\del\delbar\log|\zeta_S|^2$ to deduce that
\be\label{MJd}
J\lrcorner\delta^2(S)=-\frac1{4\pi}\Delta\log|\zeta_S|^2
\ee
where $\zeta_S$ is a section of the line bundle $\calo(S)$ that vanishes on $S$. Furthermore, since $\omega_A=\ii\del\delbar\kappa_A$ is harmonic, we can observe that 
\be\label{MJo}
J\lrcorner\omega_A=-\frac12\Delta\kappa_A
\ee
is a constant. Then, recalling (\ref{Mnorm}) and (\ref{warpeqM}), we can combine (\ref{MJd}) and (\ref{MJo})   to obtain the identity
\be
\int_Ye^{-6D_0}[J\lrcorner\delta^2(S)]\d\text{vol}_Y=\frac{1}{4\pi\ell^6_{\rm M}}\int_Y\left(2\pi m^A\kappa_A-\log|\zeta_S|^2\right)Q_8
\ee
where $2\pi m^A\kappa_A-\log|\zeta_S|^2$ is a globally defined function. 
Hence, by using (\ref{q8}), (\ref{Mwarpdiv}) finally takes the form 
\be
\begin{aligned}
\frac1{3!}\int_Se^{-6D}J\wedge J\wedge J=&\,\frac1{3!} c\,m^A\cali_{ABCD}v^Bv^Cv^D+\frac12\sum_Im^A\kappa_A(Z_I,\bar Z_I;v)\\
&\,+\frac{1}{4\pi\ell^6_{\rm M}}\int_Y\left(2\pi m^A\kappa_A-\log|\zeta_S|^2\right)\left(\frac1{2} G_4\wedge G_4-\ell^6_{\rm M}I_8\right)\\
&\,-\frac{1}{4\pi}\sum_I\left[\log\zeta_S(Z_I)+\log\bar\zeta_S(\bar Z_I)\right]\ .
\end{aligned}
\ee
By choosing $S\rightarrow S_a$ and $m^A\rightarrow m^A_a$ and discarding the (hol+$\overline{\rm hol}$) term appearing in the last line, we obtain (\ref{Mfirstcont}).


\subsection{Variation of $\Re\rho_a$ in M-theory}
\label{app:delrho}

We want to compute how $\Re\rho_a$ varies under a deformation $\delta v^a$ of the non-universal K\"ahler moduli $\delta v^a$ preserving the constraint (\ref{Mcon0}) and the cohomological primitivity condition (\ref{cohoprim}).

We first compute the variation of the potentials $\kappa_A(z,\bar z;v)$, starting from $\delta\omega_A=\ii\del\delbar\delta\kappa_A $, following \cite{Martucci:2014ska}. This can be done without restricting to deformations preserving (\ref{cohoprim}). Hence, for the moment, we consider more general variation $\delta u^A$ preserving (\ref{Mcon0}).
Defining 
\be
d_A\equiv J\lrcorner \omega_A
\ee
(which is a constant along $Y$) one can check  that we can write
\be\label{delkappa}
\Delta\delta\kappa_A=-2\left(\delta d_A+\delta u^B\omega_B\lrcorner\omega_A\right)\ .
\ee 
Noticing that 
\be
\delta d_A=\frac{1}{2{\rm w}_0}\cali_{ABCD}u^Cu^D\delta u^B=-\frac{1}{{\rm w}_0}\delta u^B\int_Y\omega_B\lrcorner \omega_A\d{\rm vol}_Y
\ee
we see that (\ref{delkappa}) is indeed integrable. The solution can be written as
\be
\delta\kappa_A(y;v)=\delta u^B\int_{Y,y'}G(y,y')(J\wedge J\wedge \omega_A\wedge \omega_B)(y')
\ee
where $G(y,y')=G(y',y)$ is the Green's function associated with the K\"ahler metric $\d s^2_Y$.

Restricting back to variations $\delta v^a$ preserving the primitivity condition (\ref{cohoprim}), we can now write
\be\label{Mdelkappa}
\delta\kappa_A(y;v)=\delta v^a\int_{Y,y'}G(y,y')(J\wedge J\wedge \omega_a\wedge \omega_A)(y')
\ee
where $\omega_a=m_a^A\omega_A$. By writing $e^{-6D_0}$ appearing in (\ref{Msplit}) as 
\be
e^{-6D_0(y)}=\frac1{\ell^8_{\rm M}}\int_Y G(y;y') Q_8
\ee
it is not difficult to check that the variation (\ref{Mdelkappa}) in (\ref{Mrho}), combined with the variation of $\frac1{3!}c\cali_{aABC}u^Au^Cu^D$,  produces 
\be\label{Mrhovar1}
\frac12\,\delta v^b\int_Y e^{-6D}J\wedge J\wedge \omega_a\wedge\omega_b\,\quad\subset\quad \delta\rho_a\ .
\ee

On the other hand, in computing $\delta\rho_a$ we also need to take into account the hidden $G_4$ dependence on the $v^a$ K\"ahler moduli discussed in subsection \ref{sec:Mkahler}. By using (\ref{G4var}) and (\ref{C2var}) it is straightforward to compute the corresponding variation in  (\ref{Mrho}). In particular the variation of  $G_4$ in $h(v)$  combines with the variation of the last term (the $\int_{S_a}$ integral), producing  
\be\label{Mrhovar2}
\frac{1}{2\ell^6_{\rm M}}\delta v^b\int_Y\omega_a\wedge\Lambda^{1,1}_b\wedge  G_4\,\quad\subset\quad \delta\rho_a\ .
\ee
By combining (\ref{Mrhovar1}) and (\ref{Mrhovar2}) one obtains (\ref{Mderho}).


\subsection{Useful formulas for M-theory effective theory}
\label{app:Meff}

Starting from the definition (\ref{Mrho}) and using (\ref{Mderho}), one can rewrite the tautological identities 
$\frac{\del\Re\rho_a}{\del\Re\rho_b}=\delta^b_a$,  $\frac{\del\Re\rho_a}{\del Z^i_I}=0$ and  $\frac{\del\Re\rho_a}{\del \beta^\alpha}=0$ as follows
\begin{subequations}\label{Minv}
\begin{align}
\frac1{3!}\cali_{aABC}u^A u^B u^C\frac{\del c}{\del\Re\rho_b}+\caln_{ac}\frac{\del v^c}{\del\Re\rho_b}&=\delta^b_a\label{Minv1}\\
\frac1{3!}\cali_{aABC}u^A u^B u^C\frac{\del c}{\del Z^i_I}+\caln_{ab}\frac{\del v^b}{\del Z^i_I}+\frac12\cala^I_{ai}&=0\label{Minv2}\\
\frac1{3!}\cali_{aABC}u^A u^B u^C\frac{\del c}{\del \beta^\alpha}+\caln_{ab}\frac{\del v^b}{\del \beta^\alpha}+\frac12\calt_{a\alpha\bar\beta}\bar\beta^{\bar\beta}&=0\label{Minv3}
\end{align}
\end{subequations}
where $\cala^I_{ai}$ is defined in (\ref{Mcala}). By contracting (\ref{Minv}) with $v^a$ and recalling (\ref{Ncontr}),  we get
\be\label{Mder1}
\frac{\del c}{\del \Re\rho_a}=\frac{1}{4{\rm w}_0}v^a\,,\quad \frac{\del c}{\del Z^i_I}=-\frac{1}{8{\rm w}_0}v^a\cala^I_{ai}\,,\quad \frac{\del c}{\del \beta^\alpha}=-\frac{1}{8{\rm w_0}}v^a\calt_{a\alpha\bar\beta}\bar\beta^{\bar\beta}\ .
\ee
By using these relations, (\ref{Minv})  also give
\be\label{Mder2}
\begin{aligned}
\frac{\del v^a}{\del\Re\rho_b}&=\caln^{ab}-\frac{1}{12{\rm w}_0 c}v^a v^b\\
\frac{\del v^a}{\del Z^i_I}&=-\frac{1}{2}\left(\caln^{ab}-\frac{1}{12{\rm w}_0 c}v^a v^b\right)\cala^I_{bi}\\
\frac{\del v^a}{\del \beta^\alpha}&=-\frac12\left(\caln^{ab}-\frac{1}{12{\rm w}_0 c}v^a v^b\right)\calt_{b\alpha\bar\beta}\bar\beta^{\bar\beta}
\end{aligned}
\ee
where $\caln^{ab}$ is the inverse of $\caln_{ab}$.


\end{appendix}




\providecommand{\href}[2]{#2}\begingroup\raggedright\endgroup


\end{document}